# Non-detection of the Tooth Fairy at Optical Wavelengths

E. Armstrong[1, 2*]

[1]*Department of Physics, University of California: San Diego, La Jolla, CA, 92037*
[2]*Department of Astronomy, Columbia University, Mail code 5246, 550 West 120th St, New York, NY 10027*

## ABSTRACT

We report a non-detection, to a limiting magnitude of V = 18.4 (9), of the elusive entity commonly described as the Tooth Fairy. We review various physical models and conclude that follow-up observations must precede an interpretation of our result.

## 1 INTRODUCTION

The Tooth Fairy is a common figure worldwide, and particularly in the United States and Europe (Wikipedia 2012). The creature is associated with a recently-lost human tooth that is usually deciduous in nature. With as-yet undetermined motivations, the creature purloins these teeth and in exchange leaves behind small gifts. The recently de-toothed person, of age 5.6 (18) (Seuss, Dr. 1981), expects this burglary and, in fact, eagerly anticipates it, typically placing the tooth in a readily accessible location[1]. Despite potential barriers such as bolted front doors and ornery Rottweilers, it appears that the Tooth Fairy obtains the tooth with minimal difficulty, arriving and departing undetected. Indeed, the only observational evidence of the being's transient presence is the vanishing of the tooth in question.

While there exists general consensus regarding the Tooth Fairy's perplexing fetish for discarded human remains, the precise identity of the Tooth Fairy is a contentious topic. Fig. 1 depicts a representative array of suggested visual counterparts. The majority (74%) believe the Tooth Fairy to be a female of approximately[2] human form (Fig. 1a), while the remaining 26% suggest a myriad of alternatives including man (Fig. 1c), bear (Fig. 1d), bat (Fig. 1g), bunny rabbit (Fig. 1f), dragon (Fig. 1b), and dental hygenist (Fig. 1e). For simplicity, we shall hereafter refer to the Tooth Fairy using the feminine gender class.

To date, all pictorial representations summarized in Fig. 1 are speculative. A robust detection of the Tooth Fairy has not been obtained at any bandwidth. A tentative infrared detection was reported based on the experience of a six-year-old Nebraskan boy: "I felt a warm breath on my ear, but it mighta just been my stupid kid sister trying to steal my iPod again" (Fox News 1997). No follow-up observation was conducted. Furthermore, the boy's kid sister subsequently awoke with a suspiciously-fresh black eye, lending credence to his theory that she had been the bedside visitor. Searches for optical and high-energy counterparts to the Tooth Fairy have not been conducted. Here we report on a study aimed to remedy this oversight.

[1]In Europe and the United States, traditionally a child places the tooth under a pillow at bedtime. The Tooth Fairy's gift subsequently appears under this pillow the following morning. In some Asian countries, tradition somewhat complicates the Tooth Fairy's job of finding the tooth: in Korea and Vietnam, for example, a child typically throws the tooth upwards if it came from the lower jaw, or down to the ground if it came from the upper jaw, with the goal of encouraging future teeth to grow in a desirable orientation. The Tooth Fairy is less common in African cultures.

[2]Embellishments include wings, fairy dust, and ostentatiously pointy ears.

* E-mail: earmstrong@physics.ucsd.edu

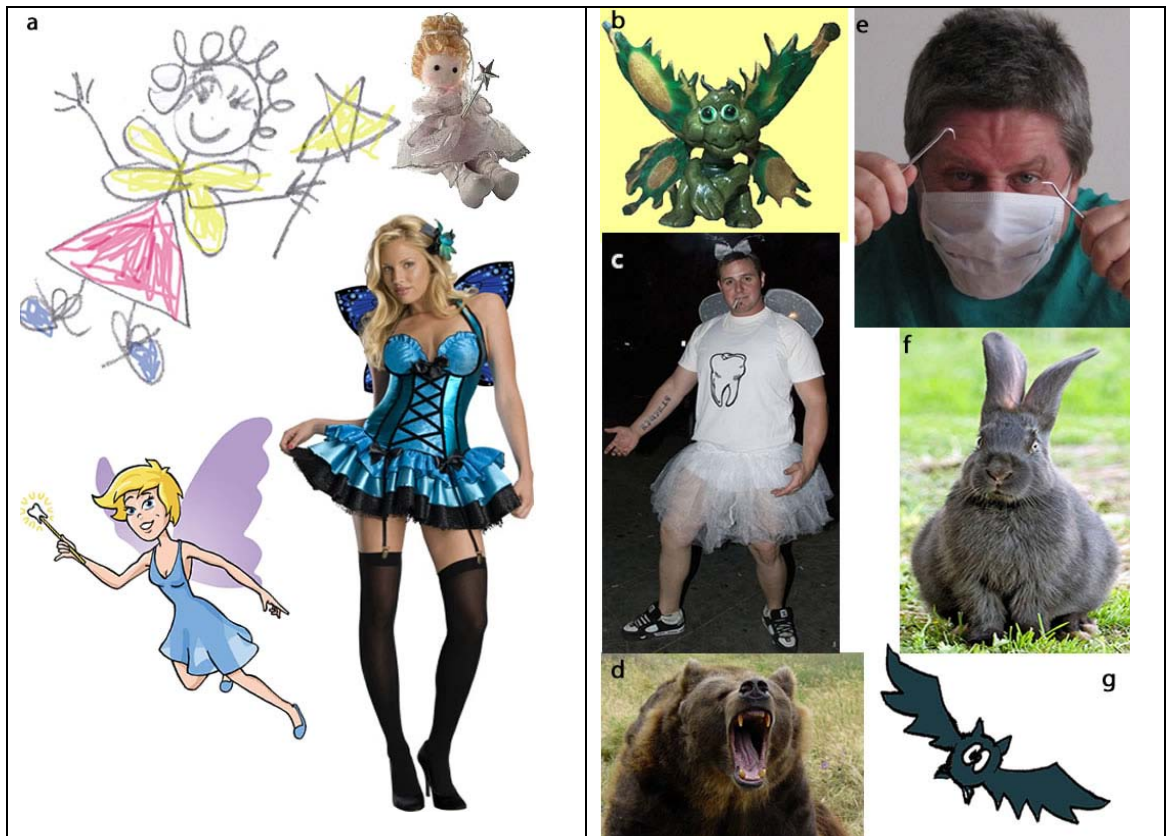

**Figure 1a**: Common depictions of the Tooth Fairy as a female of approximately-human form.    **Figure 1b-g**: Other representations.

*References.* (a: top left) Armstrong 2012a; (a: top right) overstock.com 2012; (a: bottom left) Isbister & Clotworthy 2012; (a: bottom right) Rubies' Costumes 2012; (b) Yeager 2011; (c) Sandoval 2012; (d) Johnson 2012; (e) Armstrong 2012b; (f) stock.xchng.com 2012; (g) Armstrong 2012c.

## 2   OBSERVATIONS

Observations were conducted with the MDM 1.3-meter telescope, using a CCD specially designed for near-Earth observations. The chip is 92,084 x 92,084 pixels, with a pixel length of 24 microns. The full unbinned chip yields a 0.38-arcsecond-per-pixel field of view. To push the limiting magnitude on each exposure, we set the integration time at four seconds. This time resolution suited our purposes, assuming that an open-air tooth robbery might be as fast as five seconds. In addition, we used a quarter of the chip area binned 2x2, to reduce readout time to around four seconds. The resulting limiting V magnitude on each image was 18.4 (9).

On the evening of 2012 Mar 31, this author configured a sleeping arrangement on the roof of the neighboring 2.4-meter observatory, which has an unobstructed line of sight to the 1.3m dome slit. The distance of the author's sleeping bag to the 1.3m telescope input pupil was 47 (1) m. A wisdom tooth, freshly removed from the author's lower left jaw, was placed under a pillow, upon which the author subsequently laid her head and fell asleep.

The telescope was programmed to obtain an eight-hour time series of a two-meter-radius circle centered on the author's sleeping bag. For a distance of 47 m, the limiting absolute magnitude $M_v$ is 97.5 (9).

We performed data reduction via standard UNIX and IRAF scripts. The resulting light curve is a plot of differential magnitude with respect to a 1.0 (6)-Watt flashlight that was left shining near the sleeping bag throughout the night. Fig. 2 shows the light curve of the eight-hour time series; the right panel is one representative image. Virtually all frames look alike, with the exception of a brief incident occurring around HJD = 2456018.5000 (see *Analysis*). The sky remained photometric throughout the night. Finally, at the end of the observation, the author's wisdom tooth could not be localized.

## 3   ANALYSIS

### 3.1   Features of the light curve and power spectrum

The single noteworthy characteristic of the light curve is the brief brightening at HJD = 2456018.5000-1, as shown in Fig. 2. An inspection of a fresh bite upon the author's forehead the following morning, in addition to the author's veritably-annoyed personal narrative, indicated that the optical counterpart was a mosquito that had ventured close approach to the flashlight. The average mosquito luminosity over the two images upon which it appeared was 0.005 (1) W, which yielded a differential magnitude with respect to the 1.0 (6) W flashlight of V = 5.75 (9). Otherwise the light curve appears unremarkable.

Next we created a Lomb-Scargle periodogram of the full time series. The Lomb-Scargle technique is similar to the discrete Fourier transform and is equivalent to least-squares fitting. We performed this transformation because it is what our software does and we hadn't any better ideas. The left panel of Fig. 3 shows the power spectrum, with the strongest feature around 143-9 cycles per day (c/d) noted (the low frequency resolution is not entirely unexpected, given our sample size of two points). To assess the feature's significance, we performed Monte Carlo simulations assuming a typical error of 0.02 mag. The false alarm probability is 97.2 per cent.

To search for weaker periodicities, we modeled the 143-9 c/d signal as a sinusoid, finding a best fit at $\nu$ = 145.40 c/d, and subtracted this wave from the original light curve. The prewhitened power spectrum (Fig. 3: Right) looks even worse, which is also unsurprising, given our attempt to model an unresolved feature as a wave of one discrete frequency. At this point we gave up and went inside to make breakfast.

### 3.2   The tooth

As noted above, the wisdom tooth that had been placed beneath the author's pillow prior to the start of observations was not present under the author's head the following morning. (Neither, incidentally, was the pillow. The pillow had tumbled down the sloped roof and come to alight upon a tumbleweed that was resting aside the dormitory wall). We searched for the tooth for five entire minutes, the result of which confirmed our initial null finding.

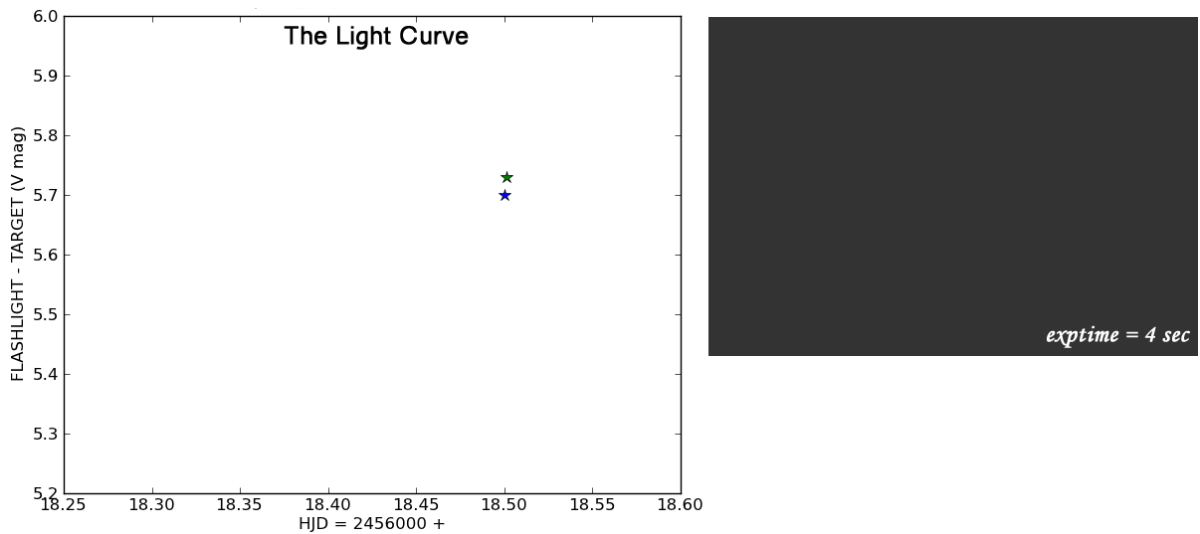

**Figure 2**: *Left*: Light curve of eight-hour time series (2012 March 31). *Right:* A representative four-second exposure from the time series.

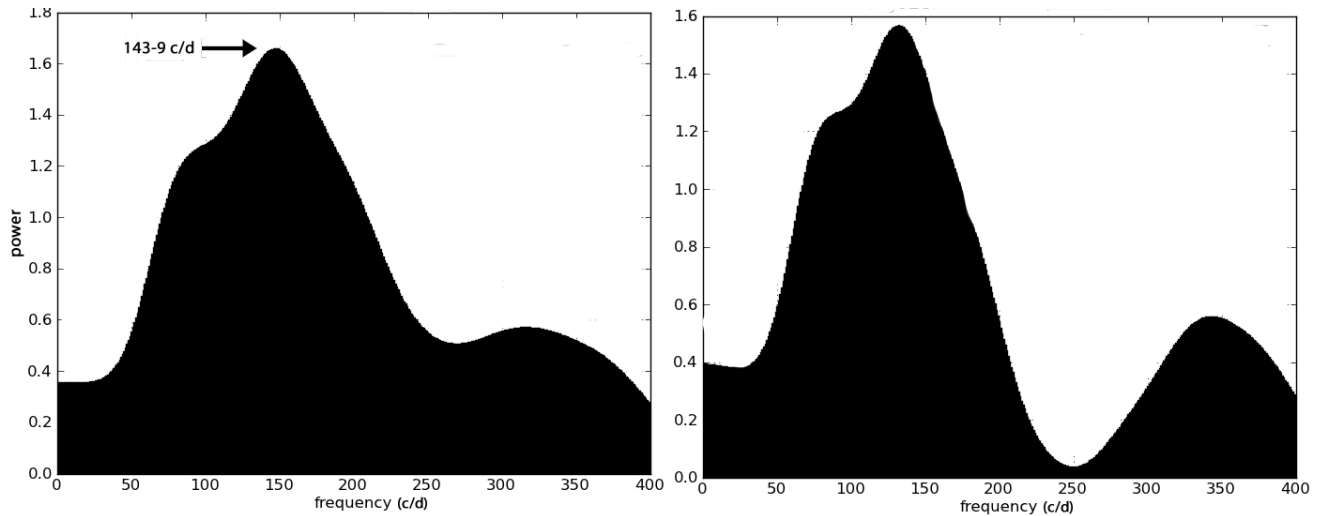

**Figure 3**: *Left*: Lomb-Scargle power spectrum of full light curve. The strong feature at ~ 145 c/d is noted. *Right*: Prewhitened power spectrum, after the attempted removal of the strongest feature (best-fit υ = 145.40 c/d).

## 4    DISCUSSION: *Faster-than-light Tooth Fairy?*

Given the tooth disappearance, we conclude that the Tooth Fairy visited during the night. How, then, to account for our non-detection? First, we remind the reader of our limiting time resolution of four seconds for both the exposures and deadtime between images. We had placed a lower limit of five seconds on an expected Tooth Fairy visit, including arrival and departure. However, this estimate is based on the typical timeframe for a *human-based* crime. The assumption that the Tooth Fairy operates on human timescales is speculative, and may have rendered erroneous the lower limit on total visit duration. Indeed, recent discoveries indicate that some non-human entities such as neutrinos (e.g. Adam et al., 2011; Ereditato 2012) and the president of Iran (The Colbert Report 2012) may indeed travel at near-luminal speeds. Thus it is desirable to obtain an observation of higher time resolution, using a collecting area sufficiently large to offset any loss of brightness sensitivity.

follow-up data in the optical band at higher time resolution with an instrument of larger collecting area, as well as multi-wavelength studies, which have yet to be conducted.

In particular, we expect optical and infrared observations to be relatively straightforward to conduct. Radio detection will present a significant creative challenge, as the surface area to be resolved is on the order of a meter. High-energy observations are probably premature: because it is not known whether the Tooth Fairy ever leaves the Earth's atmosphere, a non-detection from satellite imagery will not necessarily be diagnostic. Thus, a more thorough theoretical framework must precede an interpretation of any high-energy observations.

Finally, we note preliminary evidence that not only is the Tooth Fairy transparent at optical wavelengths, she is also a stingy sonnuvabitch. She left behind nary a penny! We tentatively attribute this swindle to spite: perhaps the creature took offense at what might be interpreted as a crass and deliberate invasion of her privacy.

## 5    CONCLUSION

We report a non-detection of the Tooth Fairy at optical wavelengths fainter than V = 18.4 (9), using a time resolution of four seconds. The distance of 47 (1) meters to our target yields a limiting absolute magnitude of $M_V$ = 97.5 (9). It is important to obtain

### ACKNOWLEDGEMENTS

This research was funded by quarters pilfered from the snack machine up at the Kitt Peak Visitors' Center.

[3]Okay, we admit it: this one made up.

Colbert (prod); Jim Hoskinson (dir). Comedy